# Supercurrent in Graphene Josephson Junctions with Narrow Trenches in the Quantum Hall Regime

Andrew Seredinski[1,*], Anne Draelos[1], Ming-Tso Wei[1], Chung-Ting Ke[1], Tate Fleming[2], Yash Mehta[2], Ethan Mancil[2], Hengming Li[2], Takashi Taniguchi[3], Kenji Watanabe[3], Seigo Tarucha[4,5], Michihisa Yamamoto[4,5], Ivan V. Borzenets[6], François Amet[2], and Gleb Finkelstein[1]

[1]Department of Physics, Duke University, Durham, NC 27708, U.S.A.
[2]Department of Physics and Astronomy, Appalachian State University, Boone, NC 28607, U.S.A.
[3]Advanced Materials Laboratory, NIMS, Tsukuba 305-0044, Japan
[4]Department of Applied Physics, University of Tokyo, Bunkyo-ku, Tokyo 1-8656, Japan
[5]Center for Emergent Matter Science (CEMS), RIKEN, Wako-shi, Saitama 351-0198, Japan
[6]Department of Physics, City University of Hong Kong, Kowloon, Hong Kong SAR
* Corresponding author: Andrew Seredinski (ams168@duke.edu)

**ABSTRACT**

Coupling superconductors to quantum Hall edge states is the subject of intense investigation as part of the ongoing search for non-abelian excitations. Our group has previously observed supercurrents of hundreds of picoamperes in graphene Josephson junctions in the quantum Hall regime. One of the explanations of this phenomenon involves the coupling of an electron edge state on one side of the junction to a hole edge state on the opposite side. In our previous samples, these states are separated by several microns. Here, a narrow trench perpendicular to the contacts creates counterpropagating quantum Hall edge channels tens of nanometres from each other. Transport measurements demonstrate a change in the low-field Fraunhofer interference pattern for trench devices and show a supercurrent in both trench and reference junctions in the quantum Hall regime. The trench junctions show no enhancement of quantum Hall supercurrent and an unexpected supercurrent periodicity with applied field, suggesting the need for further optimization of device parameters.

**INTRODUCTION**

Superconductor-quantum Hall heterostructures are predicted to host Majorana zero modes (MZMs) which could be harnessed for fault-tolerant quantum computing [1, 2, 3]. This has sparked a renewal of experimental interest in the intersection of quantum Hall physics and superconductivity [4, 5, 6, 7, 8, 9]. Recently, our group reported on supercurrent mediated by quantum Hall (QH) edge states in boron nitride (BN) encapsulated graphene [9, 10]. This follows theoretical predictions [11, 12] and raises the possibility of MZMs that could take advantage of the gate-tuneability of QH states. Exploration of higher applied magnetic fields and fractional quantum Hall states could yield parafermions: yet more exotic non-abelian anyons [13, 14]. The microscopic origin of the supercurrent in QH systems is not yet conclusively established [10, 15], and with observed critical currents on the order of hundreds of picoamps, the signature remains difficult to measure.

At zero applied magnetic field, supercurrent is mediated through the bulk of a superconductor-normal metal-superconductor Josephson junction (JJ) via Andreev bound states. These bound states emerge when an electron incident on one terminal enters the gapped superconductor as a Cooper pair and a hole is reflected back to the opposite terminal. There, a Cooper pair is annihilated and an electron is sent back across the normal region [16, 17]. The mechanism for QH supercurrent may involve a novel Andreev process in which an electron propagating along one edge of the junction is reflected along the opposite edge as a hole. The electron and hole states are predicted to be coupled via an electron-hole hybrid mode running along the width of the superconductor-normal interface [9, 12].

The magnitude of the critical current may be limited by the length of this electron-hole hybrid mode [9]. Varying the width of JJs has previously been shown to have an ambiguous impact on the strength of the supercurrent [10]. However, this study involved junction widths on the order of microns. Ideally, these edge states should be brought within one superconducting coherence length, $\xi$. On this length scale, the Andreev process involved could be analogous to the typical Andreev bound state picture. The introduction of a thin trench at the centre of the device induces a pair of closely spaced counterpropagating states. This close coupling could enhance the supercurrent carried by these edges, enabling the study of higher fields and fractional quantum Hall states.

This paper presents measurements of JJs in low magnetic fields and at $B = 1T$ with and without thinly etched trenches. The trench width is measured to be roughly 30nm, the same order of magnitude as the coherence length of the superconductor Molybdenum-Rhenium ($\xi \leq 10nm$). The existence of the trench is shown to change the Fraunhofer interference pattern at small fields, and the QH resistance at high field. However, supercurrent in the quantum Hall regime shows no enhancement and an unexpected periodicity with applied field.

**DEVICE FABRICATION**

The BN-graphene-BN heterostructure was assembled with a variation of the standard dry transfer "stamping" method [18] from mechanically exfoliated flakes of hexagonal boron nitride and Kish graphite. The "stack" was deposited on a silicon chip with a 280nm oxide layer. Monolayer graphene was identified optically before encapsulation and the single layer nature was confirmed with Raman spectroscopy [19]. The device region was identified via AFM to be free of defects and interlayer contaminants. Electron beam lithography was carried out using an FEI XL30 SEM-FEG on a sample spun coat with 180nm of PMMA.

The e-beam dose required to define thin lines in resist varies with the thickness of the underlying BN, which is tens of nanometres thick. A dose test was completed on the stack itself. To avoid damage to the device region, this was written tens of microns away. The dose test was etched using the same recipe that was used for the final trench write (see below). The resulting lines were imaged using an SEM to determine the dose yielding the thinnest continuous trench. The optimal line dose for a 30kV e-beam on this device was determined to be 2nC/cm, resulting in trenches of 30-40nm width.

Following calibration, device trenches were defined along with the usable device "mesa." The trench e-beam write created voids in the PMMA after developing that were barely detectable optically, while the write defining the mesa was easily seen. This allowed for the etching process to be tracked. Since the mesa and trench writes received the same etching procedure, the appearance of the mesa served as a good indication that the trenches cut through the graphene. First, a pure $SF_6$ reactive ion etch was used to selectively etch boron nitride [20] down to the encapsulated graphene. This was followed by an etch with $CHF_3/O_2$ (at a ratio of 10:1) which etched through both graphene and boron nitride [18].

Josephson junctions were created via sputtered Molybdenum-Rhenium (MoRe) contacts. A side view representation of one JJ of this type is shown in Figure 1a. MoRe is a type-II superconductor with a high critical temperature ($T_C \approx 10K$) and upper critical field ($H_{C2} > 9T$), which is above the magnetic field required to induce Landau quantization in the junctions.

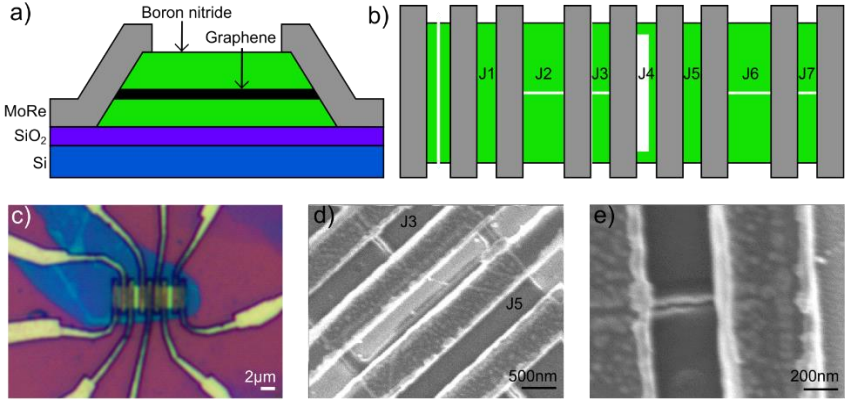

**Figure 1.** a) Side-view representation of an encapsulated graphene JJ with MoRe contacts on a $SiO_2$ / Si chip. b) Schematic view of the device showing the junction labelling convention (J1-J7). c) Optical image of the device. The form shaped around the finished device is unetched BN which was protected from the $SF_6$ step by the intervening layer of graphene and not fully etched by the $CHF_3/O_2$ process. d) SEM image of J3, J4, and J5 illustrating the presence and absence of a trench in J3 and J5 respectively. e) SEM image of J3. The trench has a width of about 30nm.

The device is shown schematically in Figure 1b and optically in Figure 1c. The Josephson junctions are labelled J1 through J7 with their dimensions summarized in Table I; width (W) is the distance along the normal-superconductor interface of the junction, while length (L) is the distance between the superconductors. J1 and J5 are reference junctions which lack trenches. J3 and J7 are copies of these but with thin trenches etched through the centre. J2 and J6 are double the length of the other junctions and also have trenches. J4 is not relevant to this study. Finally, there is a vertically cut and unnumbered junction which behaves as an open circuit, confirming the efficacy of the etch. A similar procedure has been used to confirm the cutting of graphene in the case of Helium ion milling [21, 22]. After measurement, the width of the trenches was confirmed to be 30-40nm using scanning electron microscopy. Sample scans are shown in Figure 1d-e.

**Table I.** Junction dimensions and types

| Junction | Width (W) | Length (L) | Type |
|---|---|---|---|
| J1 | 3μm | 500nm | Reference |
| J2 | 3μm | 1μm | Trench |
| J3 | 3μm | 500nm | Trench |
| J4 | 3μm | 500nm | Not studied |
| J5 | 3μm | 500nm | Reference |
| J6 | 3μm | 1μm | Trench |
| J7 | 3μm | 500nm | Trench |

## RESULTS

Measurements were conducted in a Leiden Cryogenics dilution refrigerator at a previously confirmed electron base temperature of 35mK. Details on the filtering setup of the system were previously published [23]. Standard four-terminal transport measurements were conducted using the two MoRe contacts on either side of the JJ. Each one of these contacts splits off into two leads away from the device. Backgate voltages were applied through the silicon substrate to tune the carrier density of the graphene, and magnetic field was applied perpendicular to the junction plane.

Ballisticity of the graphene is confirmed via the presence of Fabry-Pérot oscillations in gate sweeps of the resistance. Figure 2a shows normal resistance vs. gate of J7 at base temperature but current biased into the normal state. This normal state resistance approaches the ballistic limit for a junction of its dimensions. Figure 2b provides the associated bias-gate map of the voltage measured across the junction at base temperature. These demonstrate the sharp Dirac peak as well as the Fabry-Pérot oscillations of the normal resistance and supercurrent at negative gate voltages. These oscillations result from the MoRe contacts, which locally n-dope the graphene due to a work function mismatch; reflections of the ballistic carriers off these p-n interfaces result in a gate-dependent interference effect [5, 24, 25].

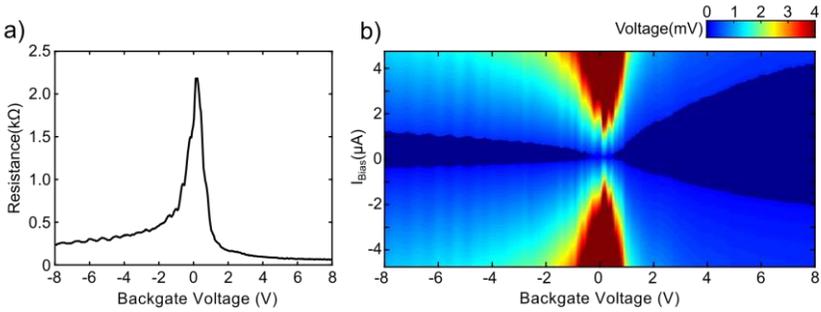

**Figure 2.** a) Normal state resistance of J7 demonstrating a sharp Dirac peak and Fabry-Pérot resistance oscillations. b) Bias-gate map of the voltage measured across the same junction at base temperature. The dark region around zero bias indicates vanishing resistance, meaning superconductivity.

## Low Field

The Fraunhofer pattern demonstrates the interference of critical current distributed across a JJ as a function of perpendicular applied magnetic field [26]. The variation of the superconducting phase parameter across the width of the junction as a function of magnetic flux results in periodic fields of vanishing critical current. These nodes appear with every magnetic flux quantum threaded through the junction, with a "sidelobe" of supercurrent in between each node. Due to flux focusing from the superconducting contacts, the effective area of the junction is increased to $W \times L + 2 \times (A/2)$ where $A$ is the area of a MoRe contact (500nm by 3µm) [9]. For the 500nm by 3µm junctions (J1, J3, J5, and J7), this yields an expected periodicity of ~0.7mT.

Figure 3a shows Fraunhofer interference data for two representative junctions in the studied device, J1, a reference, and J3, a similarly proportioned junction with a trench. For the trench case, a clear suppression of the second sidelobe can be seen. This observation is compared to a minimal numerical model plotted in Figure 3b. The model assumes a uniform sinusoidal current-phase relation across the junction, a uniform current distribution, and current travelling perpendicular to the superconducting contacts. As is clearly shown, the second sidelobe in this simulation does begin to disappear as a trench is etched down the centre of the JJ, but at a trench width of 30nm the change is minimal. The experimentally observed pattern is reproduced for trench widths of 300nm.

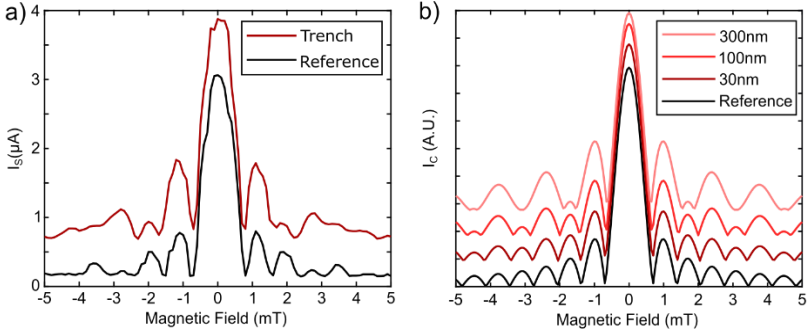

**Figure 3.** a) Fraunhofer interference patterns for J1 (reference) and J3 (trench) offset by 0.5µA. The periodicity is seen to be close to the expected 0.7mT for both junctions. The second sidelobe in the trench sample is greatly diminished. b) Numerical model of Fraunhofer interference patterns for 500nm by 3µm junctions with trenches of different widths, offset and plotted in arbitrary units of current for this qualitative comparison. The disappearance of the second sidelobe for wider trenches and the expected periodicity are clearly shown.

This feature is not surprising: as the centre of the junction is removed and the trench width is increased, the Fraunhofer pattern should smoothly morph into a SQUID-like interference pattern indicative of two interfering paths. This manifests in part as a disappearance of the even-numbered sidelobes.

The order of magnitude difference in the trench width showing this effect is indicative of the simplicity of the model. First, the current carrying modes across the junction can be thought of as having a finite width on the order of the Fermi wavelength ($\lambda_F$) at a given charge density [27]. Calculated from the applied backgate voltage (10V) using a parallel plate capacitor model, $\lambda_F \approx 40nm$. If the trench disrupts the nearest mode on either side, its effective width more than triples for the purposes of this measurement. The model also neglects particle trajectories that are not directly perpendicular to the contacts. An outsized percentage of these trajectories will be affected by the trench.

This modelling is nevertheless valuable as it demonstrates that the Fraunhofer pattern is expected to be sensitive to the presence of trenches, as is seen experimentally. It should be noted that the model plots critical current ($I_C$), while the measurement is of switching current ($I_S$); this comparison is not rendered spurious as the qualitative behaviour of $I_S$ does not differ from that of $I_C$ as a function of magnetic field.

## **High Field**

When a perpendicular field creates an electron cyclotron radius smaller than half the smallest junction dimension ($r_c < 0.5 \times \min(W, L)$), Landau quantization ensues and the quantum Hall effect is observed. The evolution of QH plateaus in a JJ configuration differs from that of a standard Hall bar, as JJs are necessarily two-terminal devices; the shape of the steps between plateaus depends on the aspect ratio of the junction [28]. This is observed in Figure 4a-b where QH plateaus at an applied field of 1T are shown for J1 and J3. Additionally, the quantization on the n-doped side of the gate sweeps is stronger than on the p-side, a result of the MoRe contacts locally n-doping the graphene.

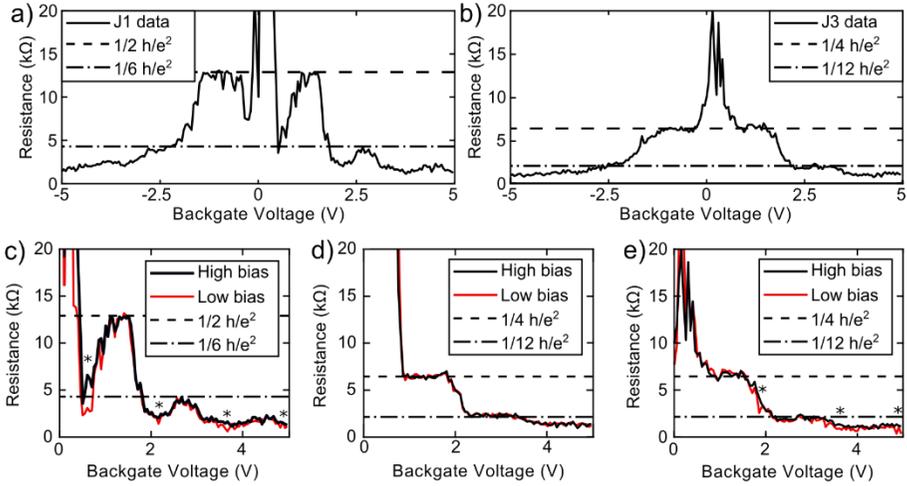

**Figure 4.** a)-b) Gate sweeps of J1 and J3 demonstrating well-defined Hall resistance plateaus at a field of one tesla. The relevant fractions of the resistance quantum $h/e^2$ are also marked, with a) giving $\nu = 2$ and $\nu = 6$, and b) showing half of these values due to the trench. c)-e) High and low bias gate sweeps of J1-J3 in the QH regime ($B = 1T$). Supercurrent appears as dips in the low bias data relative to the high bias data, marked by asterisks above the curves. While pockets are clearly visible in J1 and J3, there is no supercurrent in J2 due to the doubled junction length.

It is immediately clear that, while the reference J1 follows a typical filling factor pattern for graphene ($\nu = 2, 6, 10, ...$), the plateau resistances for the trench junctions are halved. This is the result of the trench cutting the junction into two parallel resistors, each with an identical quantized Hall resistance. This demonstrates that the trench yields QH states on either side as expected.

QH supercurrent is demonstrated by measuring with a small (tens of picoamps) AC signal while sweeping gate voltage in two conditions: once with a "high" DC bias of a few nanoamps, and once with a "low" (zero) DC bias [9]. The magnitude of the QH supercurrent is typically below 1nA, meaning that a bias current of a few nanoamps is enough to suppress it, resulting in an enhanced differential resistance. The two gate sweeps can then be compared, and areas of lower resistance in the zero DC bias condition are identified as "pockets" of supercurrent. This is seen in Figure 4c-e with regions of supercurrent marked with asterisks. These high and low DC bias gate sweeps of J1, J2, and J3 demonstrate QH supercurrent in only J1 and J3. The supercurrent is absent in J2 at this field due to the doubled length of the junction [10]. The gate location of QH supercurrent pockets is not well understood but is known to vary with applied field [9].

The supercurrent magnitude can be determined via a bias-gate map, as shown in Figure 5a-b. The superconducting branch is observed as a dip in resistance on a plateau. The differential resistance of these maps is plotted in log scale to make the resistance dips visually apparent. Here, it is demonstrated that the supercurrent magnitude at one tesla is not significantly different for J1 and J3. It should be noted that the zero-field switching and retrapping currents of these junctions are comparable, meaning that the contact transparency and cleanness of these junctions is similar.

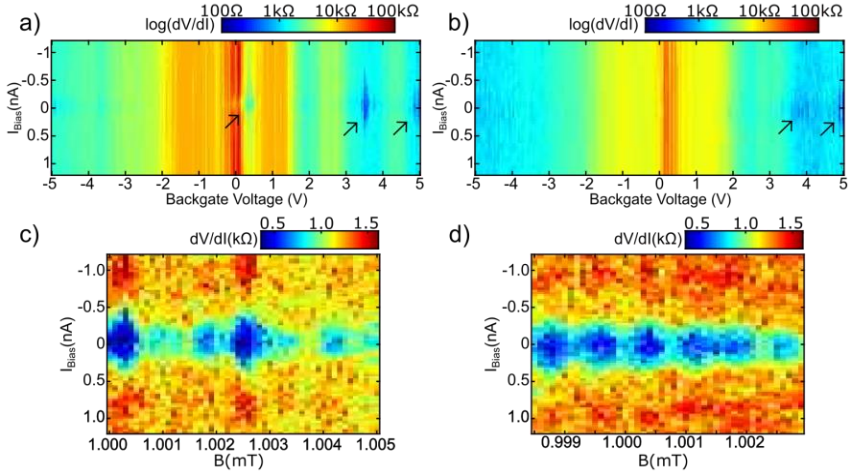

**Figure 5.** a)-b) Bias-gate maps of J1 and J3 in applied fields of $B = 1T$ demonstrating pockets of supercurrent (arrows) on and between quantized Hall plateaus. c) Interference pattern in magnetic field of a supercurrent pocket in reference junction J1 (3.56V in backgate, second arrow in 5a) showing a weak 0.7mT periodicity. d) The same measurement for a pocket in trench junction J3 (4.96V in backgate, second arrow in 5b) also demonstrating a 0.7mT periodicity.

The periodicity of these supercurrent pockets is important to explore, as it has ramifications for the possible mechanism of QH supercurrent transport. While the supercurrent periodicity in applied field is predicted to be h/e [11, 12], previous studies have demonstrated h/2e periodicity [9, 10] matching the low-field Fraunhofer pattern. This is borne out in this experiment as well. Figure 3 shows periodicities in low magnetic field on the order of 0.7mT in both reference and trench junctions, consistent with h/2e given flux focusing and the dimensions of the junctions. A very similar period is also observed around 1T, shown in Figure 5c-d.

Surprisingly, there is no change in the periodicity of the supercurrent in the trench junctions, as is seen in Figure 5d. Geometrically, a doubling of the periodicity is expected, as the QH states now encircle regions of roughly half the previous area. It is known that the trenches did yield QH states on either side due to the halved QH resistance plateaus. This unintuitive periodicity is consistent with a picture in which only the most distant QH states are involved in supercurrent transfer. Further study is required to understand this result.

**CONCLUSIONS**

This investigation demonstrates formation of trenches of 30nm width in encapsulated graphene Josephson junctions which generate changes in the Fraunhofer interference pattern and a doubling of the quantum Hall conductance. However, these trenches do not lead to QH supercurrent enhancement. The periodicity of the supercurrent at one tesla is consistent with previous results [9, 10] and does not change with the introduction of a trench.

Further optimization will lead to the etching of thinner trenches into these heterostructures. Future work will utilize higher spin speeds and lower dilutions of PMMA to achieve thinner resists and improve feature resolution. The etching process can also be optimized by minimizing the CHF$_3$/O$_2$ step to destroy the exposed graphene and limiting unwanted widening of the trench.

**ACKNOWLEDGEMENTS**

A.S., A.D., M.T.W., C.T.K., and G.F. were supported by ARO Award W911NF-16-1-0122 and NSF awards ECCS-1610213 and DMR-1743907. T.F., Y.M., E.M., H.L., and F.A. acknowledge the ARO under Award W911NF-16-1-0132. I.V.B. acknowledges CityU New Research Initiatives/Infrastructure Support from Central (APRC): 9610395. M.Y. acknowledges support from the Canon Foundation and JSPS KAKENHI Grant Number JP25107003. K.W. and T.T. acknowledge support from JSPS KAKENHI Grant Number JP15K21722 and the Elemental Strategy Initiative conducted by the MEXT, Japan. T.T. acknowledges support from JSPS Grant-in-Aid for Scientific Research A (No. 26248061) and JSPS Innovative Areas "Nano Informatics" (No. 25106006). S.T. acknowledges support from JSPS Grant-in-Aid for Scientific Research S (No. 26220710) and A (No. 16H02204). This work was performed in part at the Duke University Shared Materials Instrumentation Facility (SMIF), a member of the North Carolina Research Triangle Nanotechnology Network (RTNN), which is supported by the National Science Foundation (Grant ECCS-1542015) as part of the National Nanotechnology Coordinated Infrastructure (NNCI).

The authors acknowledge Dr. Sergey Yarmolenko of North Carolina A&T University for assistance with Raman spectroscopy for sample characterization.